\pdfoutput=1
\documentclass[preprint,preprintnumbers,amsmath,amssymb,floatfix,endfloats*]{revtex4}
\usepackage{graphicx}% Include figure files
\usepackage{dcolumn}% Align table columns on decimal point
\usepackage{bm}% bold math
\usepackage{amsfonts}

\DeclareMathAlphabet{\mathsfsl}{OT1}{cmr}{bx}{it}
\begin{document}
%----------------------------------------------------------------------%
% Title
%----------------------------------------------------------------------%
\title{The influence of complex thermal treatment on mechanical properties of amorphous materials}
\author{Qing-Long Liu$^{1}$ and Nikolai V. Priezjev$^{1,2}$}
\affiliation{$^{1}$National Research University Higher School of
Economics, Moscow 101000, Russia}
\affiliation{$^{2}$Department of Mechanical and Materials
Engineering, Wright State University, Dayton, OH 45435}
\date{\today}
\begin{abstract}

We study the effect of periodic, spatially uniform temperature
variation on mechanical properties and structural relaxation of
amorphous alloys using molecular dynamics simulations. The
disordered material is modeled via a non-additive binary mixture,
which is annealed from the liquid to the glassy state with various
cooling rates and then either aged at constant temperature or
subjected to thermal treatment. We found that in comparison to aged
samples, thermal cycling with respect to a reference temperature of
approximately half the glass transition temperature leads to more
relaxed states with lower levels of potential energy.  The largest
energy decrease was observed for rapidly quenched glasses cycled
with the thermal amplitude slightly smaller than the reference
temperature. Following the thermal treatment, the mechanical
properties were probed via uniaxial tensile strain at the reference
temperature and constant pressure.  The numerical results indicate
an inverse correlation between the levels of potential energy and
values of the elastic modulus and yield stress as a function of the
thermal amplitude.

\vskip 0.5in


\end{abstract}

\maketitle

\section{Introduction}

Due to their disordered structure, metallic glasses are known to
possess high strength and elastic limit as well as high resistance
to corrosion, which makes them potentially suitable for various
structural and biomedical applications~\cite{Egami13,ZhengBio16}.
These amorphous alloys, however, suffer from lack of ductility and
typically exhibit brittle fracture due to shear band formation,
especially in aged samples~\cite{Lewandowski05}. Commonly used
methods to rejuvenate glasses and to improve plasticity, such as
shot-peening~\cite{Concustell09}, cold rolling and
drawing~\cite{Eckert11}, as well as high-pressure
torsion~\cite{Joo15}, typically cause severe plastic deformation. By
contrast, it was recently shown that structural rejuvenation in
metallic glasses can be induced by temporarily heating samples above
$T_g$ and subsequently quenching with a suitably fast cooling rate
for recovery annealing~\cite{Ogata15,Ogata17}.  It was later found
that for some alloys, thermal rejuvenation can be enhanced by
applying external pressure~\cite{Ogata16,Yang18}. Alternatively,
rejuvenated states in metallic glasses can be accessed via cryogenic
thermal cycling below $T_g$, which promotes local structural
transformations due to spatially non-uniform thermal
expansion~\cite{Ketov15,Greer19}.   The influence of thermal
expansion heterogeneity on rejuvenation of metallic glasses was
predicted to be important at sufficiently large length scales as
compared to scales accessible to atomistic
simulations~\cite{Barrat18}.  However, the microscopic details of
the thermal processing as well as the degree of rejuvenation or
relaxation that can be achieved by applying multiple cycles remain
to be clarified.

\vskip 0.05in

During the last decade, the dynamic response of amorphous materials
to periodic shear deformation was extensively studied using
atomistic simulations and experimental
measurements~\cite{Lacks04,Priezjev13,Sastry13,Reichhardt13,
Priezjev14,IdoNature15,Priezjev16,Priezjev16a,Sastry17,
Priezjev17,Buttinoni17,Priezjev18,Priezjev18a,Rogers18,NVP18strload}.
Notably, it was shown that thermal aging process in rapidly quenched
glasses is facilitated by repetitive subyield cycling that leads to
progressively lower levels of potential energy, and the effect is
more pronounced at larger strain amplitudes~\cite{Priezjev18,
Priezjev18a,NVP18strload}.  More recently, it was found that relaxed
states can be attained by repeatedly heating and cooling binary
glasses at constant pressure with various thermal amplitudes below
the glass transition~\cite{Priez18tcyc,Priez18T5000}. In particular,
the results of numerical simulations have shown that the largest
decrease of the potential energy and the increase in the yield
stress occur for rapidly cooled glasses with the thermal amplitude
not far below the glass transition temperature~\cite{Priez18tcyc}.
It was later found that after hundreds of thermal cycles with
respect to a very low reference temperature, the glasses evolve into
steady states, where particle dynamics becomes nearly reversible
after each cycle, similar to the so-called limit cycles observed
during athermal periodic shear of amorphous
materials~\cite{Priez18T5000,Reichhardt13,IdoNature15}. However, the
dependence of the potential energy and mechanical properties on the
preparation history, period and number of thermal oscillations,
thermal amplitude, and reference temperature remains unexplored.

\vskip 0.05in

In this paper, molecular dynamics simulations are performed to
examine the effect of thermal cycling on potential energy states and
mechanical properties of binary glasses. The thermal oscillations
with relatively small period are imposed with respect to a reference
temperature of about half the glass transition temperature. It will
be shown that regardless of preparation history, the thermal
treatment of one hundred cycles always leads to relaxed states in a
wide range of thermal amplitudes.   Subsequent tensile loading of
aged and thermally cycled glasses reveals that the yield stress and
the elastic modulus acquire maxima at the thermal amplitude that
corresponds to the most relaxed states.

\vskip 0.05in

The paper is structured as follows. The description of molecular
dynamics simulations and thermomechanical processing protocols are
provided in the next section. The analysis of the potential energy
series, particle displacements, and mechanical properties are
presented in section\,\ref{sec:Results}.  Brief conclusions are
given in the last section.

\section{Molecular dynamics simulations}
\label{sec:MD_Model}

The metallic glass is represented by the Kob and Andersen (KA)
binary mixture model, which was originally developed to study the
amorphous metal alloy
$\text{Ni}_{80}\text{P}_{20}$~\cite{KobAnd95,Weber85}. In this
model, there are two types of atoms, $A$ and $B$, with strongly
non-additive cross interactions that prevent crystallization upon
cooling below the glass transition temperature~\cite{KobAnd95}. More
specifically, the interaction between two atoms $\alpha,\beta=A,B$
is described by the truncated Lennard-Jones (LJ) potential:
\begin{equation}
V_{\alpha\beta}(r)=4\,\varepsilon_{\alpha\beta}\,\Big[\Big(\frac{\sigma_{\alpha\beta}}{r}\Big)^{12}\!-
\Big(\frac{\sigma_{\alpha\beta}}{r}\Big)^{6}\,\Big],
\label{Eq:LJ_KA}
\end{equation}
with the following parametrization $\varepsilon_{AA}=1.0$,
$\varepsilon_{AB}=1.5$, $\varepsilon_{BB}=0.5$, $\sigma_{AB}=0.8$,
$\sigma_{BB}=0.88$, and $m_{A}=m_{B}$~\cite{KobAnd95}. The LJ
potential is truncated at the cutoff radius
$r_{c,\,\alpha\beta}=2.5\,\sigma_{\alpha\beta}$ to alleviate the
computational burden. The system is composed of $48000$ $A$ atoms
and $12000$ $B$ atoms, with the total number of atoms of $60000$. In
what follows, all physical quantities are reported in terms of the
reduced LJ units of length, mass, energy, and time:
$\sigma=\sigma_{AA}$, $m=m_{A}$, $\varepsilon=\varepsilon_{AA}$, and
$\tau=\sigma\sqrt{m/\varepsilon}$, respectively. The MD simulations
were carried out using the LAMMPS parallel code with the time step
$\triangle t_{MD}=0.005\,\tau$~\cite{Lammps}.

\vskip 0.05in

% equilibration and temperature protocol

\begin{figure}[t]
\includegraphics[width=9.0cm,angle=0]{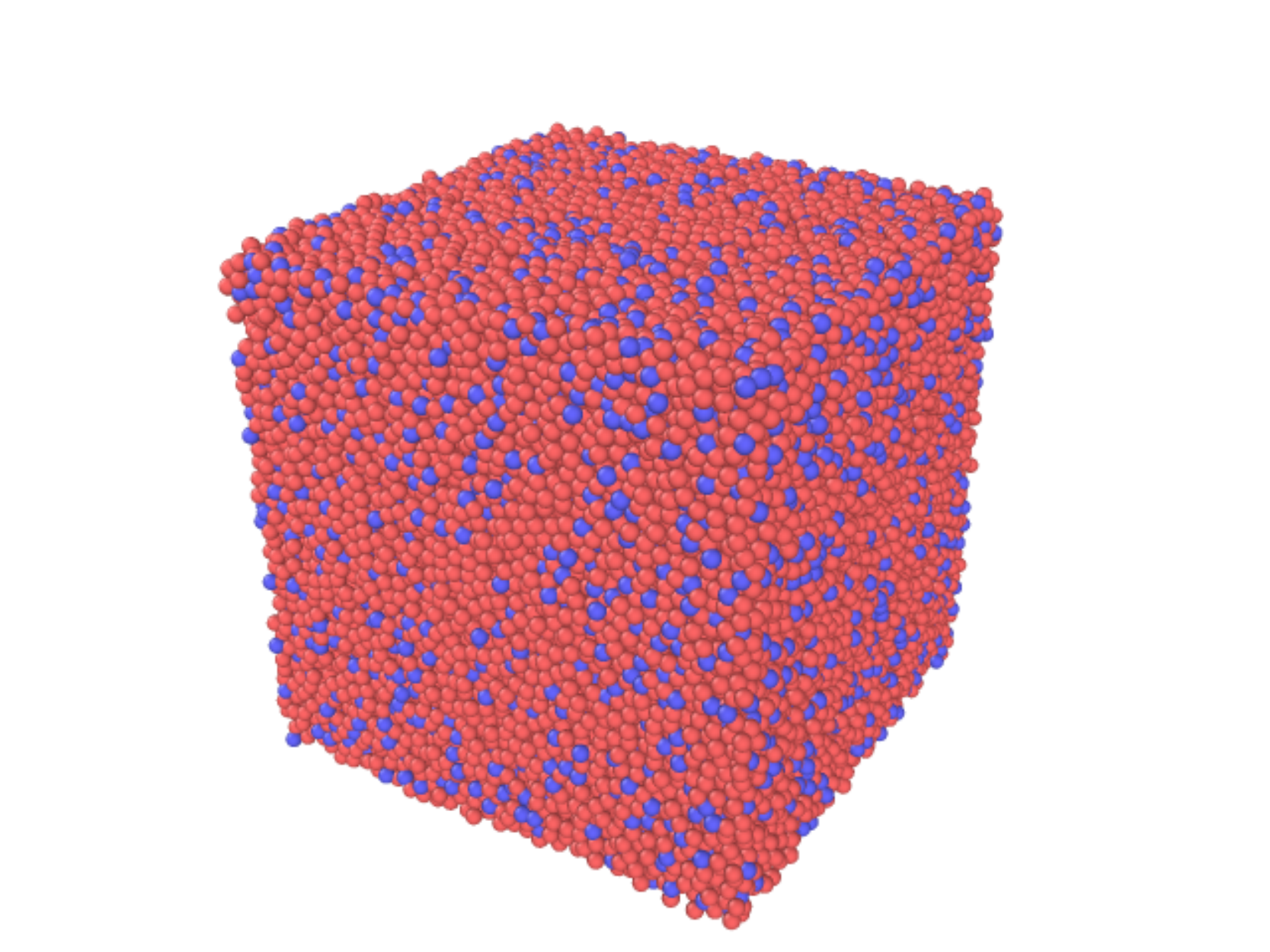}
\caption{(Color online) A representative snapshot of the binary LJ
glass annealed with the cooling rate of
$10^{-2}\varepsilon/k_{B}\tau$ to the temperature
$T_{LJ}=0.2\,\varepsilon/k_B$. The total number of atoms is
$60\,000$. The larger atoms of type $A$ are denoted by red spheres
and smaller atoms of type $B$ are indicated by blue spheres. }
\label{fig:snapshot_system}
\end{figure}

The simulations were performed in several stages. The system was
first thoroughly equilibrated at the temperature of
$0.7\,\varepsilon/k_B$, which is above the glass transition of the
KA model, $T_g\approx0.435\,\varepsilon/k_B$~\cite{KobAnd95}. Here,
$k_B$ denotes the Boltzmann constant. The temperature is controlled
via the Nos\'{e}-Hoover thermostat, and the periodic boundary
conditions are imposed along all three dimensions~\cite{Allen87}.
After the equilibration procedure, the samples were annealed at
constant pressure to the temperature $T_{LJ}=0.2\,\varepsilon/k_B$
with the cooling rates $10^{-2}\varepsilon/k_{B}\tau$,
$10^{-3}\varepsilon/k_{B}\tau$, $10^{-4}\varepsilon/k_{B}\tau$, and
$10^{-5}\varepsilon/k_{B}\tau$. The snapshot of the binary glass
annealed with the cooling rate of $10^{-2}\varepsilon/k_{B}\tau$ to
the temperature of $T_{LJ}=0.2\,\varepsilon/k_B$ is illustrated in
Fig.\,\ref{fig:snapshot_system}.  Next, the system was repeatedly
heated and cooled with the thermal amplitude $\Delta T_{LJ}$ during
100 cycles with the period $T=2000\,\tau$.   The thermal
oscillations were imposed at constant pressure, $P=0$, with respect
to the reference temperature $T_{LJ}=0.2\,\varepsilon/k_B$. After
the thermal treatment, the samples were strained along the $\hat{x}$
direction at constant pressure with the strain rate
$\dot{\varepsilon}_{xx}=10^{-5}\,\tau^{-1}$.  At each stage, the
potential energy, pressure components, system dimensions, and atomic
configurations were periodically saved for the post-processing
analysis.

\section{Results}
\label{sec:Results}

% potential energy and density for aging at T=0.2

\begin{figure}[t]
\includegraphics[width=12.0cm,angle=0]{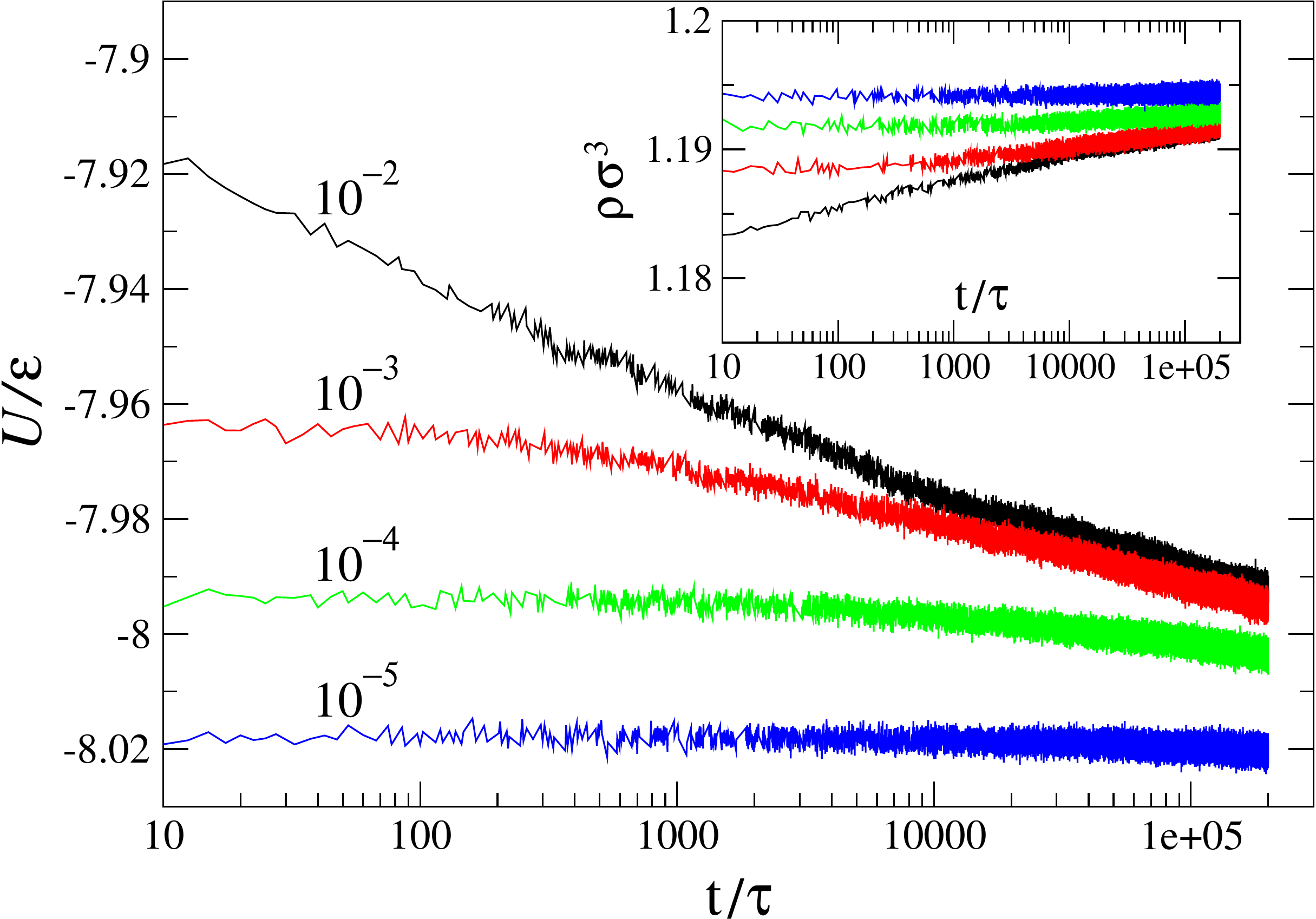}
\caption{(Color online) The variation of the potential energy per
atom for binary glasses prepared with cooling rates
$10^{-2}\varepsilon/k_{B}\tau$ (black),
$10^{-3}\varepsilon/k_{B}\tau$ (red), $10^{-4}\varepsilon/k_{B}\tau$
(green), and $10^{-5}\varepsilon/k_{B}\tau$ (blue). The simulations
are performed at constant temperature $T_{LJ}=0.2\,\varepsilon/k_B$
and zero pressure. The inset shows the glass density as a function
of time for the same samples. }
\label{fig:poten_dens_T0.2}
\end{figure}

As discussed in Sec.\,\ref{sec:MD_Model}, the binary mixture was
initially equilibrated at the temperature of $0.7\,\varepsilon/k_B$
and zero pressure, and then annealed below the glass transition
point to the temperature of $0.2\,\varepsilon/k_B$ with different
cooling rates. The starting temperature of $0.7\,\varepsilon/k_B$
was chosen to be not far above the glass transition temperature
$T_g\approx0.435\,\varepsilon/k_B$ in order to reduce the annealing
time during slow cooling and to avoid significant deformation of the
simulation domain from a cubic box, since all system dimensions were
allowed to vary independently at constant pressure. After the
glasses were annealed with the cooling rates
$10^{-2}\varepsilon/k_{B}\tau$, $10^{-3}\varepsilon/k_{B}\tau$,
$10^{-4}\varepsilon/k_{B}\tau$, and $10^{-5}\varepsilon/k_{B}\tau$,
the simulations proceeded at $T_{LJ}=0.2\,\varepsilon/k_B$ and $P=0$
during the time interval $2\times10^5\tau$.  The results for the
potential energy per atom and the average glass density during the
aging process are reported in Fig.\,\ref{fig:poten_dens_T0.2} for
the indicated values of the cooling rate.

\vskip 0.05in

% potential energy and density for aging at T=0.2; cont

It can be observed from Fig.\,\ref{fig:poten_dens_T0.2} that upon
cooling with slower rates, the potential energy levels become
deeper, as the system visits a larger number of minima in the
potential energy landscape in the vicinity of the glass transition
temperature. During the aging process at
$T_{LJ}=0.2\,\varepsilon/k_B$ and $P=0$, shown in
Fig.\,\ref{fig:poten_dens_T0.2}, the potential energy of more
rapidly quenched glasses decays significantly during the time
interval of $2\times10^5\tau$.  A similar decrease of the potential
energy during aging below $T_g$ at constant volume during $10^5\tau$
was reported in the previous MD study, where it was also shown that
the system dynamics, as measured by the the decay of the two-time
intermediate scattering function, becomes progressively
slower~\cite{KobBar97}. We also comment that the potential energy
levels reported in the recent study with a similar setup, except
that the reference temperature is $T_{LJ}=0.01\,\varepsilon/k_B$,
remained nearly constant during the time interval $10^6\tau$ and
relatively low, \textit{i.e.},
$U\lesssim-8.24\,\varepsilon$~\cite{Priez18tcyc}. As shown in the
inset to Fig.\,\ref{fig:poten_dens_T0.2}, the rate of density
increase is more pronounced for rapidly cooled glasses, and the
difference in the average glass densities after $2\times10^5\tau$ is
less than $0.3\,\%$.

\vskip 0.05in

% temperature protocol

\begin{figure}[t]
\includegraphics[width=12.0cm,angle=0]{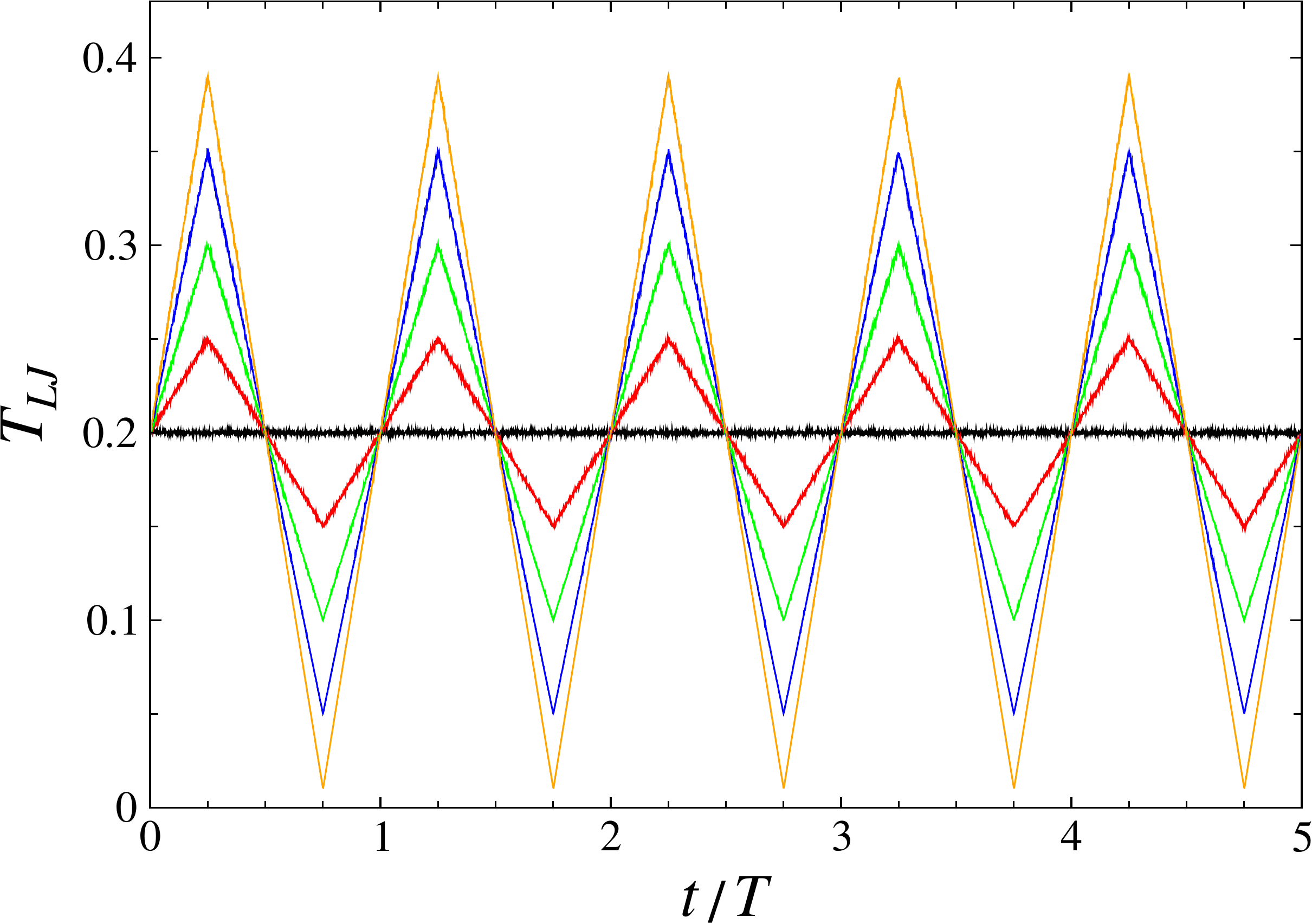}
\caption{(Color online) The variation of temperature $T_{LJ}$ (in
units of $\varepsilon/k_B$) during 5 periods, $T=2000\,\tau$, for
the thermal amplitudes $\Delta T_{LJ}=0.05\,\varepsilon/k_B$ (red),
$0.10\,\varepsilon/k_B$ (green), $0.15\,\varepsilon/k_B$ (blue), and
$0.19\,\varepsilon/k_B$ (orange). The black line denotes the data at
the constant temperature $T_{LJ}=0.2\,\varepsilon/k_B$. The data are
taken in the binary glass initially annealed with the cooling rate
of $10^{-3}\varepsilon/k_{B}\tau$ to the temperature
$T_{LJ}=0.2\,\varepsilon/k_B$. }
\label{fig:temp_control}
\end{figure}

Following the annealing procedure with different cooling rates,
$10^{-2}\varepsilon/k_{B}\tau$, $10^{-3}\varepsilon/k_{B}\tau$,
$10^{-4}\varepsilon/k_{B}\tau$, and $10^{-5}\varepsilon/k_{B}\tau$,
the glasses were subjected to repeated cycles of heating and cooling
with respect to the reference temperature
$T_{LJ}=0.2\,\varepsilon/k_B$.  The examples of the selected
temperature profiles measured during the first five cycles are
presented in Fig.\,\ref{fig:temp_control} for the glass that was
initially annealed with the cooling rate of
$10^{-3}\varepsilon/k_{B}\tau$. In what follows, we denote the
maximum deviation from the reference temperature
$T_{LJ}=0.2\,\varepsilon/k_B$, or the thermal amplitude, by $\Delta
T_{LJ}$.  In the present study, we considered a wide range of
thermal amplitudes but ensured that temperature remained above zero
and below the glass transition temperature. Thus, the maximum value
of the thermal amplitude is $\Delta T_{LJ}=0.19\,\varepsilon/k_B$.
We also remind that the simulations were carried out at constant
pressure ($P=0$), thus allowing significant variation in volume
during each cycle.   Here, we emphasize the key differences in the
choice of parameters from the previous MD study on thermally cycled
binary glasses~\cite{Priez18tcyc}. Specifically, the thermal
treatment was performed with a smaller oscillation period,
$T=2000\,\tau$, higher reference temperature,
$T_{LJ}=0.2\,\varepsilon/k_B$, and larger number of thermal
amplitudes to resolve more accurately the neighborhood around the
minimum of the potential energy after 100 cycles (discussed below).

\vskip 0.05in

% potential energy for thermal cycling

\begin{figure}[t]
\includegraphics[width=12.0cm,angle=0]{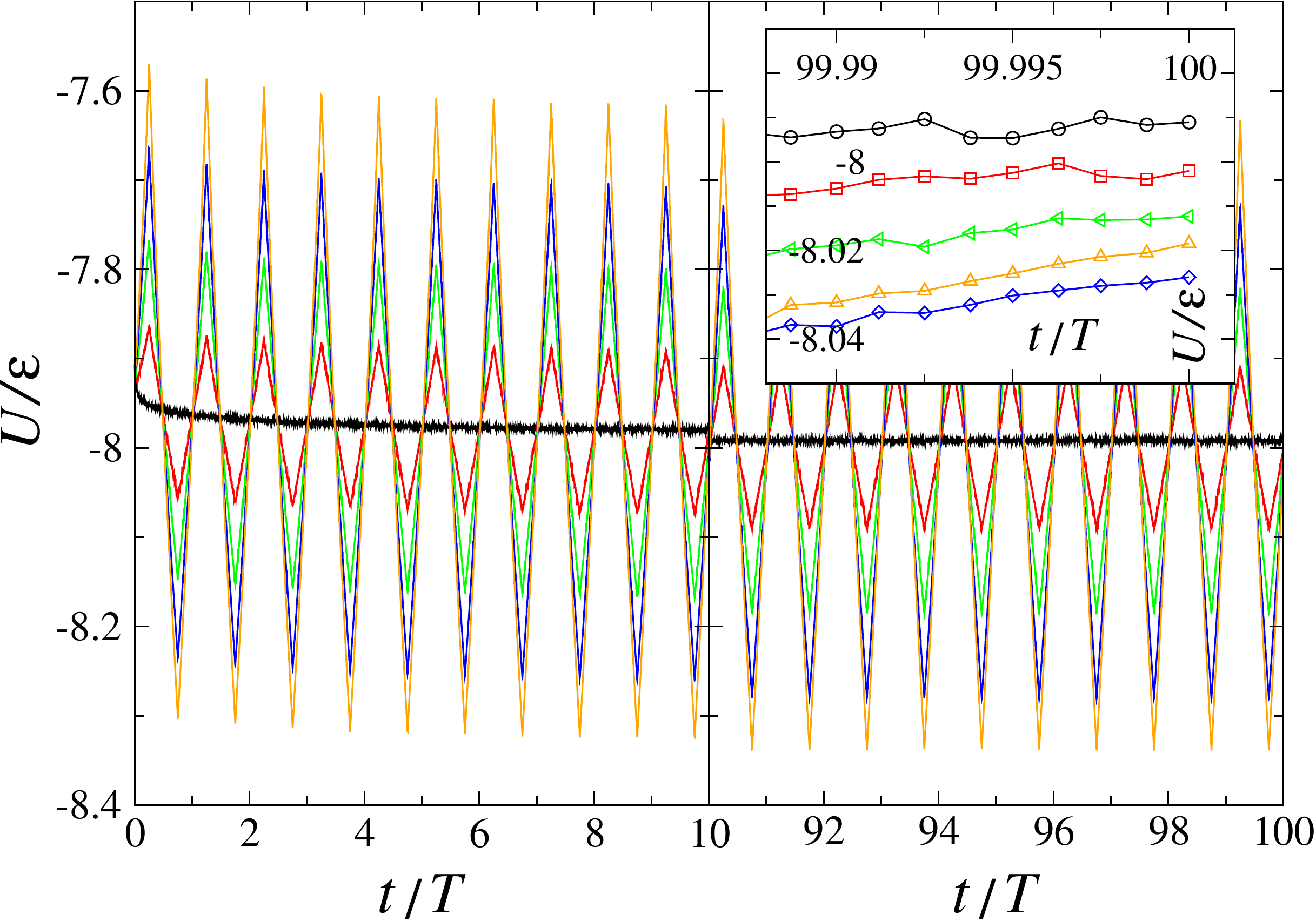}
\caption{(Color online) The potential energy series during the first
and last ten cycles with the thermal amplitudes $\Delta T_{LJ}=0$
(black), $0.05\,\varepsilon/k_B$ (red), $0.1\,\varepsilon/k_B$
(green), $0.15\,\varepsilon/k_B$ (blue), and $0.19\,\varepsilon/k_B$
(orange). The sample was initially annealed with the cooling rate of
$10^{-2}\varepsilon/k_{B}\tau$. The enlarged view of the same data
at the end of the last cycle is displayed in the inset.}
\label{fig:poten_10m2}
\end{figure}

\begin{figure}[t]
\includegraphics[width=12.0cm,angle=0]{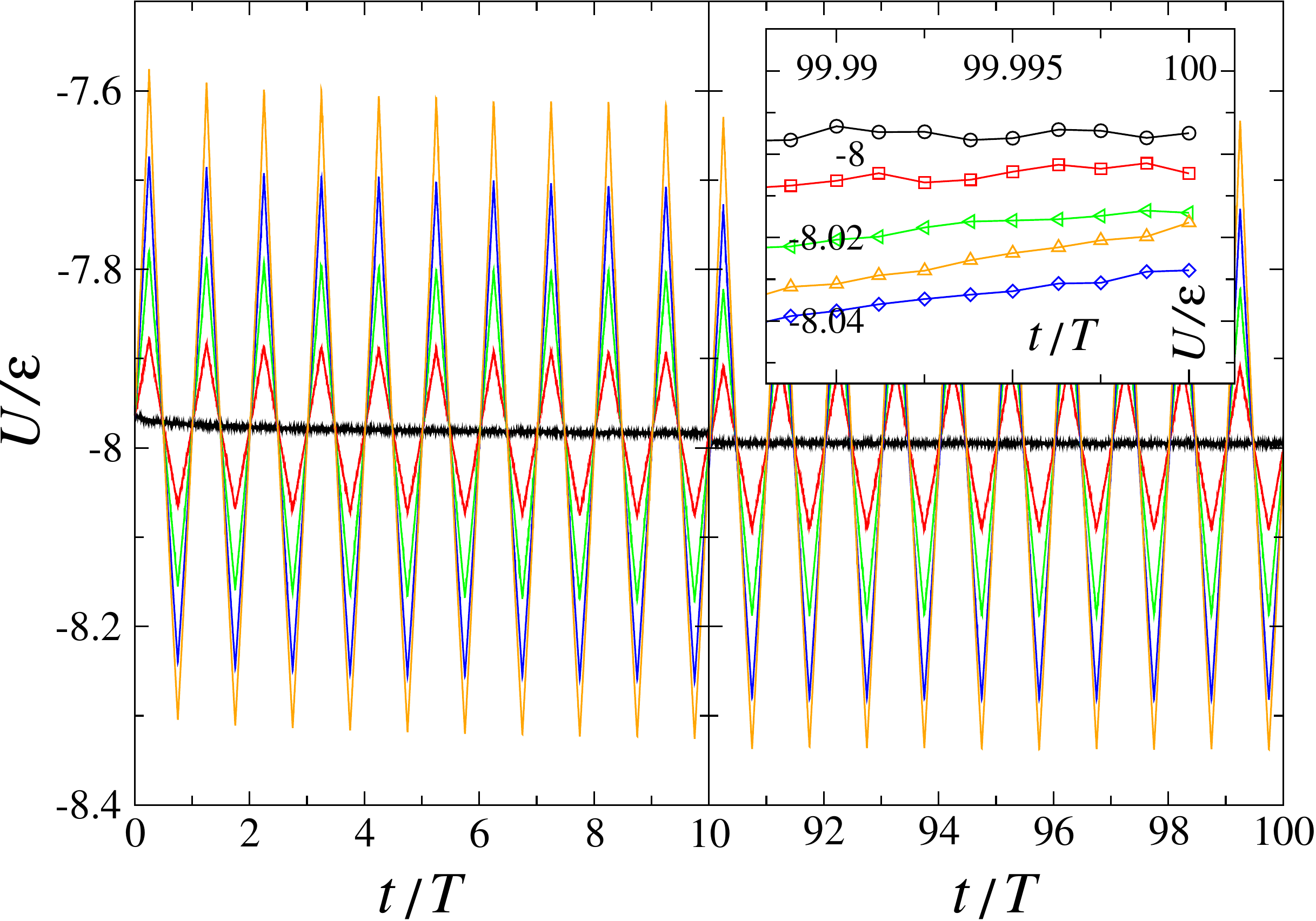}
\caption{(Color online) The variation of the potential energy for
the glass annealed with the cooling rate of
$10^{-3}\varepsilon/k_{B}\tau$ and subjected to thermal cycling with
the amplitudes $\Delta T_{LJ}=0$ (black), $0.05\,\varepsilon/k_B$
(red), $0.1\,\varepsilon/k_B$ (green), $0.15\,\varepsilon/k_B$
(blue), and $0.19\,\varepsilon/k_B$ (orange). The inset shows $U$ at
$t\approx100\,T$. }
\label{fig:poten_10m3}
\end{figure}

\begin{figure}[t]
\includegraphics[width=12.0cm,angle=0]{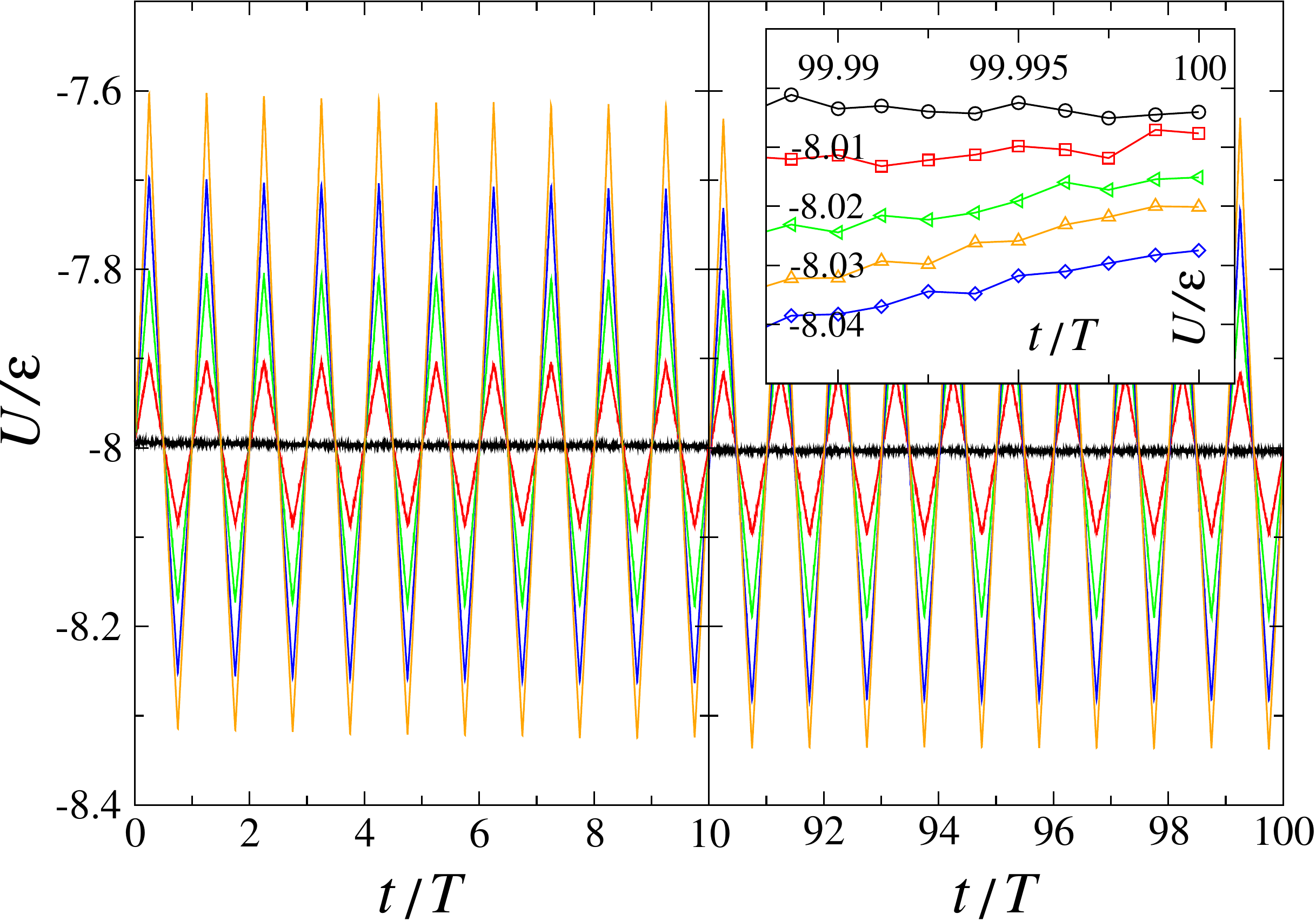}
\caption{(Color online) The potential energy per atom,
$U/\varepsilon$, for glasses cycled with the amplitudes $\Delta
T_{LJ}=0$ (black), $0.05\,\varepsilon/k_B$ (red),
$0.1\,\varepsilon/k_B$ (green), $0.15\,\varepsilon/k_B$ (blue), and
$0.19\,\varepsilon/k_B$ (orange). The cooling rate is
$10^{-4}\varepsilon/k_{B}\tau$. The same data are resolved near
$t\approx100\,T$ and shown in the inset. }
\label{fig:poten_10m4}
\end{figure}

\begin{figure}[t]
\includegraphics[width=12.0cm,angle=0]{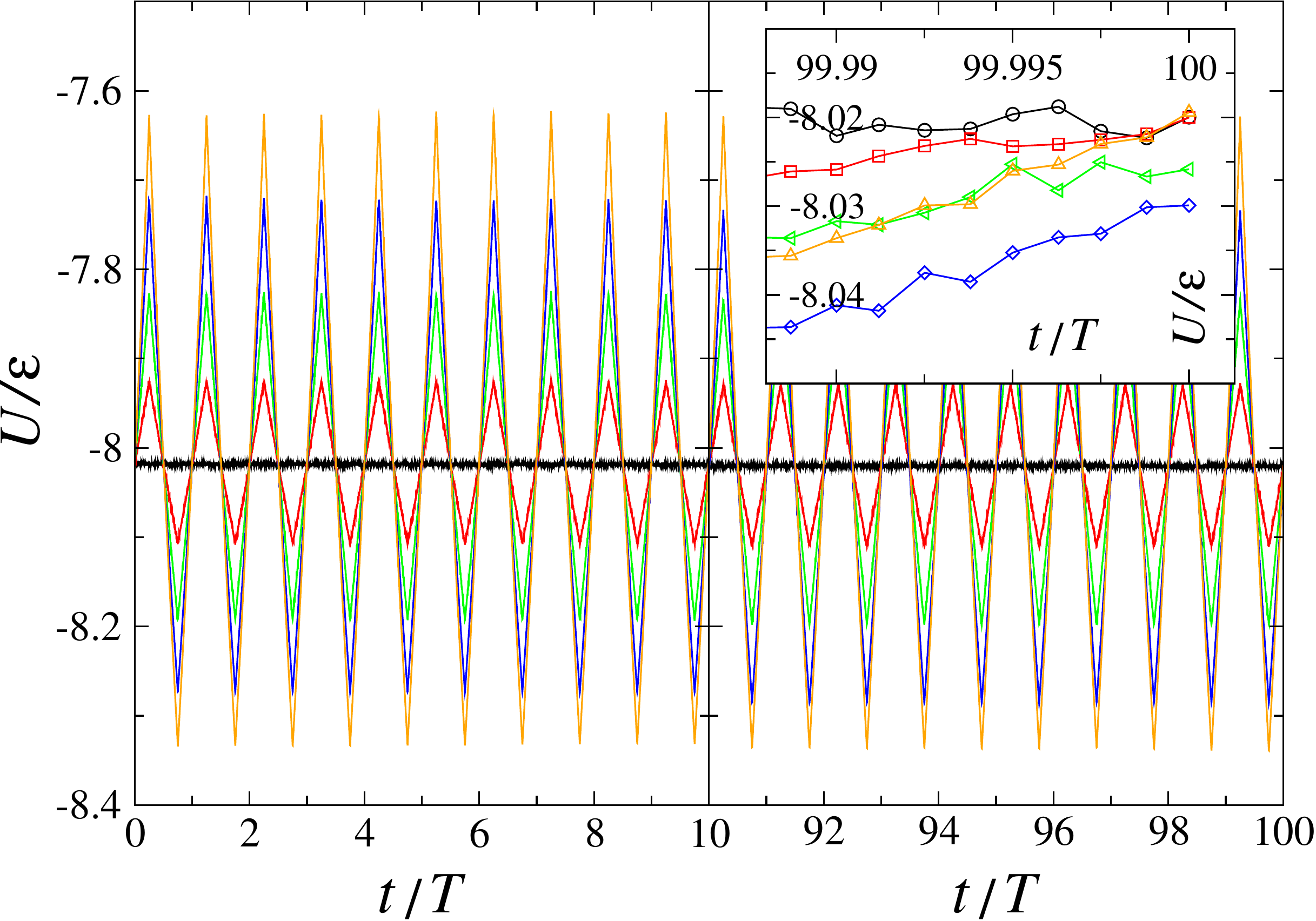}
\caption{(Color online) The time dependence of the potential energy
during thermal treatment with amplitudes $\Delta T_{LJ}=0$ (black),
$0.05\,\varepsilon/k_B$ (red), $0.1\,\varepsilon/k_B$ (green),
$0.15\,\varepsilon/k_B$ (blue), and $0.19\,\varepsilon/k_B$
(orange). The sample was initially cooled with the rate
$10^{-5}\varepsilon/k_{B}\tau$. The inset shows the same data in the
vicinity of $t\approx100\,T$.}
\label{fig:poten_10m5}
\end{figure}

We next present the variation of the potential energy at the
beginning and the end of thermal treatment in
Figs.\,\ref{fig:poten_10m2}, \ref{fig:poten_10m3},
\ref{fig:poten_10m4}, and \ref{fig:poten_10m5} for glasses initially
annealed with the cooling rates $10^{-2}\varepsilon/k_{B}\tau$,
$10^{-3}\varepsilon/k_{B}\tau$, $10^{-4}\varepsilon/k_{B}\tau$, and
$10^{-5}\varepsilon/k_{B}\tau$, respectively. For reference, the
black curves in each figure denote the data at the constant
temperature $T_{LJ}=0.2\,\varepsilon/k_B$ (the same data as in
Fig.\,\ref{fig:poten_dens_T0.2}). It can be seen that the amplitude
of the potential energy oscillations increases at larger thermal
amplitudes.  From Figs.\,\ref{fig:poten_10m2}-\ref{fig:poten_10m4},
it is apparent that for relatively quickly annealed glasses, the
minima of the potential energy become progressively deeper over
consecutive cycles for all thermal amplitudes. Moreover, a small
difference in the potential energy between aged and thermally cycled
glasses is developed after each cycle.  This discrepancy becomes
especially evident in the enlarged view of the data during the last
cycle shown in the insets to
Figs.\,\ref{fig:poten_10m2}-\ref{fig:poten_10m4}. By contrast, in
the case of slowly annealed glass, shown in
Fig.\,\ref{fig:poten_10m5}, it is difficult to visually detect any
changes in the minima of the potential energy from cycle to cycle
from the main panels. However, a more detailed view of the data in
the inset to Fig.\,\ref{fig:poten_10m5} reveals that there is a
noticeable deviation in the potential energy after the last cycle
for the thermal amplitudes $\Delta T_{LJ}=0.10\,\varepsilon/k_B$ and
$0.15\,\varepsilon/k_B$. This implies a nonmonotonic dependence of
$U(100\,T)$ as a function of $\Delta T_{LJ}$.

\vskip 0.05in

% potential energy for thermal cycling; cont

\begin{figure}[t]
\includegraphics[width=12.0cm,angle=0]{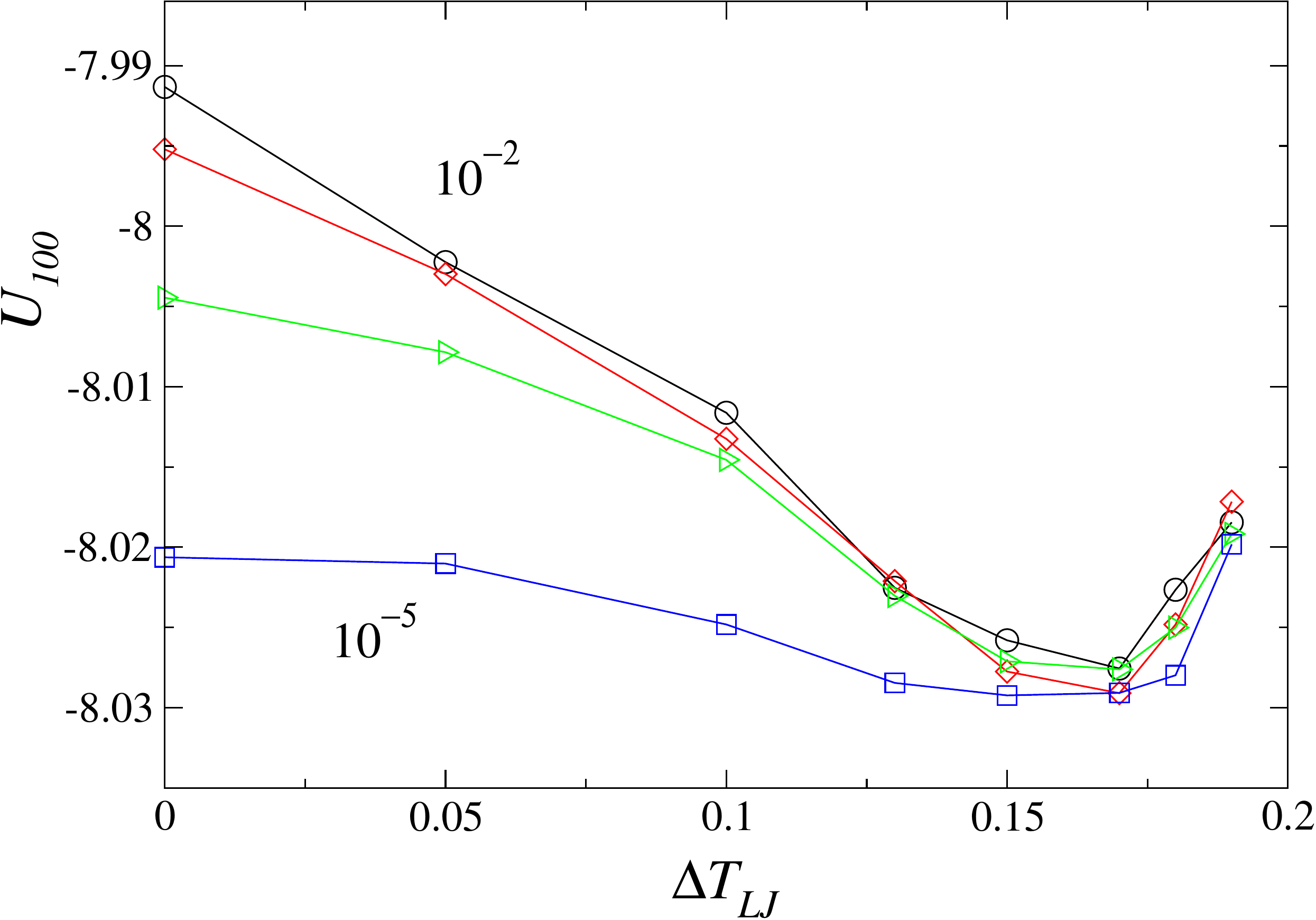}
\caption{(Color online) The dependence of the potential energy after
100 cycles, $U_{100}/\varepsilon$, as a function of the thermal
amplitude $\Delta T_{LJ}$ (in units of $\varepsilon/k_B$) for
glasses initially annealed with the cooling rates of
$10^{-2}\varepsilon/k_{B}\tau$ (black),
$10^{-3}\varepsilon/k_{B}\tau$ (red), $10^{-4}\varepsilon/k_{B}\tau$
(green), and $10^{-5}\varepsilon/k_{B}\tau$ (blue). }
\label{fig:U100_delT}
\end{figure}

We next summarize the data for the potential energy after 100
thermal cycles in Fig.\,\ref{fig:U100_delT} as a function of the
thermal amplitude for the cooling rates
$10^{-2}\varepsilon/k_{B}\tau$, $10^{-3}\varepsilon/k_{B}\tau$,
$10^{-4}\varepsilon/k_{B}\tau$, and $10^{-5}\varepsilon/k_{B}\tau$.
The data at $\Delta T_{LJ}=0$ correspond to the potential energy of
glasses aged during the time interval $2\times10^5\tau=100\,T$ at
$T_{LJ}=0.2\,\varepsilon/k_B$. The data points in
Fig.\,\ref{fig:U100_delT} were obtained by linearly extrapolating
the potential energy values in the vicinity of $100\,T$. It is
evident that with increasing thermal amplitude, the potential energy
first decreases and then acquires a local minimum at about $\Delta
T_{LJ}=0.17\,\varepsilon/k_B$. From the data presented in
Fig.\,\ref{fig:U100_delT}, it is difficult to conclude with
certainty whether the minimum in $U(100\,T)$ versus $\Delta T_{LJ}$
depends on the cooling rate.   Notice also that the dependence of
$U(100\,T)$ is nearly the same for glasses initially annealed with
cooling rates $10^{-2}\varepsilon/k_{B}\tau$ and
$10^{-3}\varepsilon/k_{B}\tau$.

\vskip 0.05in

% potential energy for thermal cycling; cont

The increase of the potential energy $U(100\,T)$ at $\Delta
T_{LJ}=0.18\,\varepsilon/k_B$ and $0.19\,\varepsilon/k_B$ in
Fig.\,\ref{fig:U100_delT} can be ascribed to relative proximity of
the system temperature after a quarter of a cycle,
$0.2\,\varepsilon/k_B+\Delta T_{LJ}$, to the glass transition
temperature.  In the latter case, the temperature approaches $T_g$
from below, the role of thermal fluctuations temporarily increases,
and the systems is then annealed with the effective cooling rate $4
\Delta T_{LJ}/T=0.00038\,\varepsilon/k_{B}\tau$ during each cycle.
Note that the potential energy of the glass annealed with the
cooling rate $10^{-5}\varepsilon/k_{B}\tau$ and aged at
$T_{LJ}=0.2\,\varepsilon/k_B$ is nearly the same as $U(100\,T)$ at
$\Delta T_{LJ}=0.19\,\varepsilon/k_B$.  A qualitatively similar
trend in the dependence of $U(100\,T)$ on $\Delta T_{LJ}$ was
reported in the recent study, although the decrease in potential
energy due to aging was not so pronounced, since the reference
temperature was much lower, \textit{i.e.},
$T_{LJ}=0.01\,\varepsilon/k_B$~\cite{Priez18tcyc}.  Overall, it can
be concluded from Fig.\,\ref{fig:U100_delT} that the sequence of 100
thermal cycles with the reference temperature
$T_{LJ}=0.2\,\varepsilon/k_B$ did not grant access to rejuvenated
states and resulted only in relaxed states for all cooling rates and
thermal amplitudes considered in the present study.

\vskip 0.05in

% displacements of particles over consecutive cycles

\begin{figure}[t]
\includegraphics[width=12.0cm,angle=0]{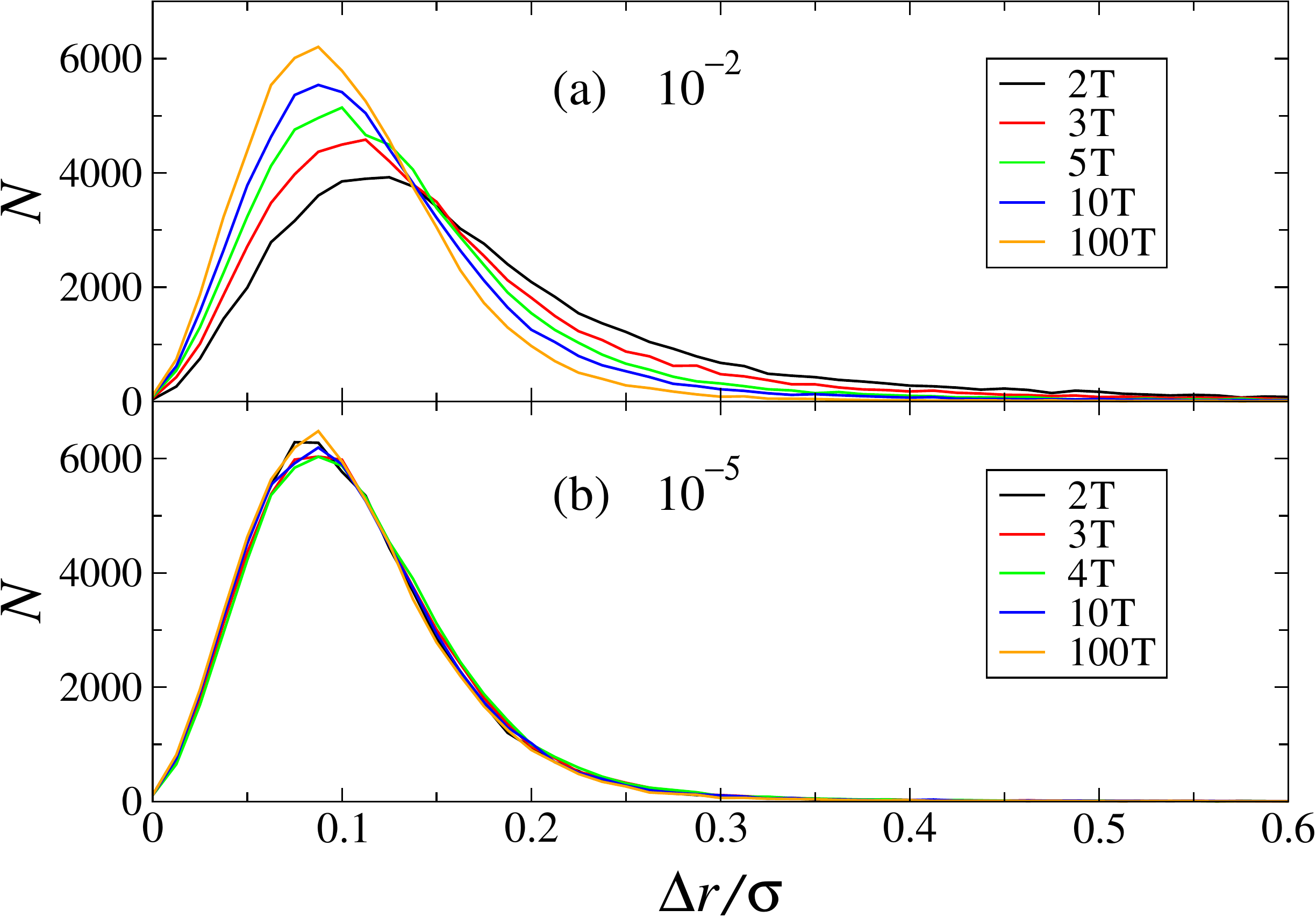}
\caption{(Color online) The distribution of atomic displacements
during one cycle for the thermal amplitude $\Delta
T_{LJ}=0.10\,\varepsilon/k_B$. The cycle numbers for computing the
displacements are tabulated in the legend.  The data are reported
for glasses initially annealed with cooling rates (a)
$10^{-2}\varepsilon/k_{B}\tau$ and (b)
$10^{-5}\varepsilon/k_{B}\tau$. The period of thermal oscillations
is $T=2000\,\tau$. }
\label{fig:PDF_TLJ01_nT}
\end{figure}

The analysis of atomic displacements during consecutive cycles
revealed that the relaxation process proceeds via irreversible
rearrangements of groups of atoms. We first examine the distribution
of displacements during selected cycles for thermal cycling with the
amplitude $\Delta T_{LJ}=0.10\,\varepsilon/k_B$. The results are
presented in Fig.\,\ref{fig:PDF_TLJ01_nT} for binary glasses
annealed with cooling rates $10^{-2}\varepsilon/k_{B}\tau$ and
$10^{-5}\varepsilon/k_{B}\tau$. It can be observed from
Fig.\,\ref{fig:PDF_TLJ01_nT}\,(a) that in the rapidly cooled glass,
the distribution of displacements is relatively broad during the
first few cycles, implying that a large number of atoms with
displacements much greater than the cage size,
$r_c\approx0.1\,\sigma$, undergo irreversible rearrangements, or
cage jumps, after one cycle. However, after 100 thermal cycles, the
distribution function becomes more narrow, indicating progressively
more reversible particle dynamics. By contrast, as evident from
Fig.\,\ref{fig:PDF_TLJ01_nT}\,(b), the shape of the distribution of
displacements for slowly annealed glass is rather insensitive to the
cycle number, and only a relatively small number of atoms might
change their cages during one cycle.

\vskip 0.05in

% displacements of particles over consecutive cycles; cont

\begin{figure}[t]
\includegraphics[width=12.0cm,angle=0]{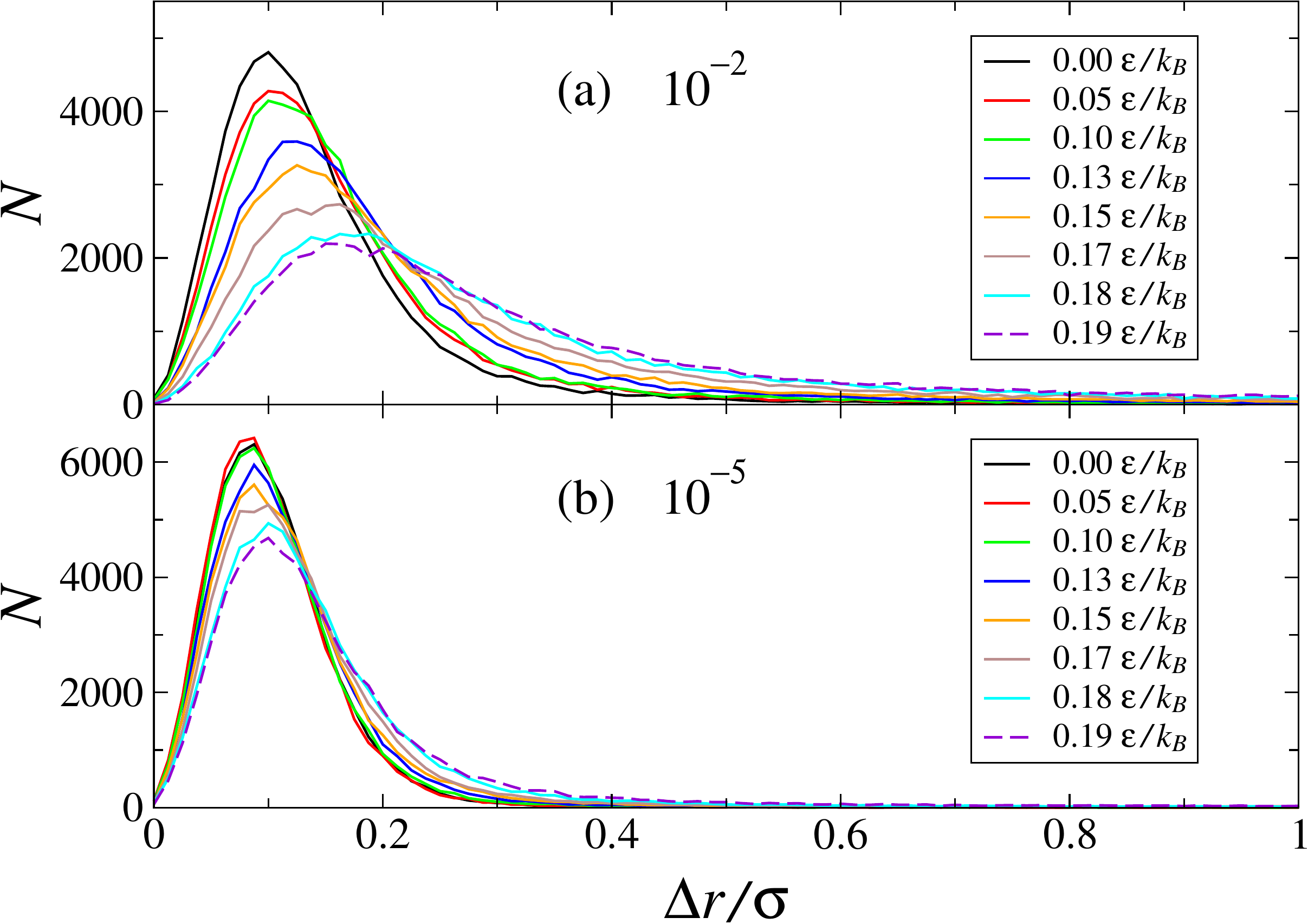}
\caption{(Color online) The probability distribution of atomic
displacements during the second cycle for the indicated values of
the thermal amplitude $\Delta T_{LJ}$.  The samples were initially
prepared with cooling rates (a) $10^{-2}\varepsilon/k_{B}\tau$ and
(b) $10^{-5}\varepsilon/k_{B}\tau$. }
\label{fig:PDF_TLJ_T}
\end{figure}

We next discuss the distribution of atomic displacements at the
beginning of the thermal treatment, i.e., during the second cycle,
but consider instead the effect of the thermal amplitude. The
probability distributions are plotted in Fig.\,\ref{fig:PDF_TLJ_T}
for selected values of the thermal amplitude and cooling rates
$10^{-2}\varepsilon/k_{B}\tau$ and $10^{-5}\varepsilon/k_{B}\tau$.
As shown in Fig.\,\ref{fig:PDF_TLJ_T}\,(a), the distribution of
displacements is much broader at the thermal amplitude $\Delta
T_{LJ}=0.19\,\varepsilon/k_B$, and it narrows upon decreasing
$\Delta T_{LJ}$ towards the limiting case of aging at constant
temperature.  A similar trend can be observed for the slowly cooled
glass in Fig.\,\ref{fig:PDF_TLJ_T}\,(b) but the effect is less
pronounced.   In general, these results are consistent with the
analysis of nonaffine displacements during thermal cycling with
respect to a very low reference temperature reported in the previous
study~\cite{Priez18tcyc}.  In particular, it was shown that atoms
with large nonaffine displacements after one cycle are organized
into transient clusters, whose size is reduced with increasing cycle
number or decreasing cooling rate and thermal
amplitude~\cite{Priez18tcyc}.

\vskip 0.05in

% mechanical properties

\begin{figure}[t]
\includegraphics[width=12.0cm,angle=0]{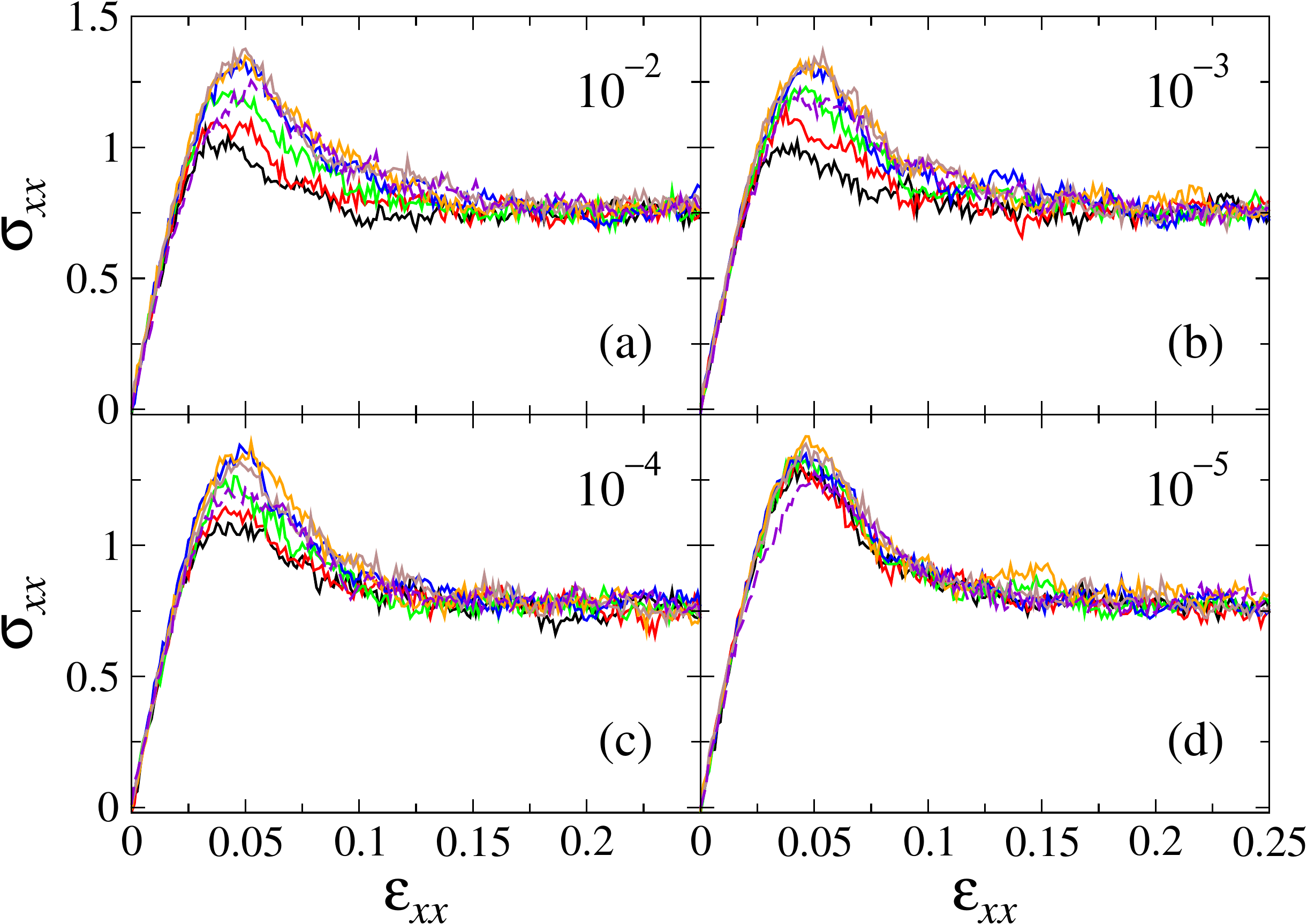}
\caption{(Color online) The variation of tensile stress,
$\sigma_{xx}$ (in units of $\varepsilon\sigma^{-3}$), as a function
of strain, $\varepsilon_{xx}$, for thermally treated glasses that
were initially annealed with cooling rates (a)
$10^{-2}\varepsilon/k_{B}\tau$, (b) $10^{-3}\varepsilon/k_{B}\tau$,
(c) $10^{-4}\varepsilon/k_{B}\tau$, and (d)
$10^{-5}\varepsilon/k_{B}\tau$. The strain rate is
$\dot{\varepsilon}_{xx}=10^{-5}\,\tau^{-1}$.  The tensile tests were
performed after the thermal treatment with amplitudes $\Delta
T_{LJ}=0$ (black), $0.05\,\varepsilon/k_B$ (red),
$0.10\,\varepsilon/k_B$ (green), $0.13\,\varepsilon/k_B$ (blue),
$0.15\,\varepsilon/k_B$ (orange), $0.17\,\varepsilon/k_B$ (brown),
and $0.19\,\varepsilon/k_B$ (dashed violet).}
\label{fig:stress_strain}
\end{figure}

After the thermal treatment, the samples were strained along the
$\hat{x}$ direction with constant strain rate
$\dot{\varepsilon}_{xx}=10^{-5}\,\tau^{-1}$ at
$T_{LJ}=0.2\,\varepsilon/k_B$ and zero pressure. The resulting
stress-strain response for different cooling rates is presented in
Fig.\,\ref{fig:stress_strain}.  For reference, the data for tensile
stress as a function of strain for glasses aged at
$T_{LJ}=0.2\,\varepsilon/k_B$ during $2\times10^5\tau$ are also
included in Fig.\,\ref{fig:stress_strain} and indicated by black
curves.  All samples were strained up to $\varepsilon_{xx}=0.25$
until the tensile stress is saturated to a constant value
independent of the processing history.   Following the elastic
regime of deformation, the stress-strain curves exhibit a pronounced
yielding peak at about $\varepsilon_{xx}=0.05$.  It can be observed
in each panel in Fig.\,\ref{fig:stress_strain} that the magnitude of
the peak becomes larger with increasing thermal amplitude up to
$\Delta T_{LJ} \approx 0.15\,\varepsilon/k_B$. Notice, however, that
the stress overshoot is generally reduced from the maximum value
when $\Delta T_{LJ} = 0.19\,\varepsilon/k_B$ for all cooling rates.
These results indicate a nonmonotonic variation of the yield stress
on the thermal amplitude.

\vskip 0.05in

% mechanical properties; cont

\begin{figure}[t]
\includegraphics[width=12.cm,angle=0]{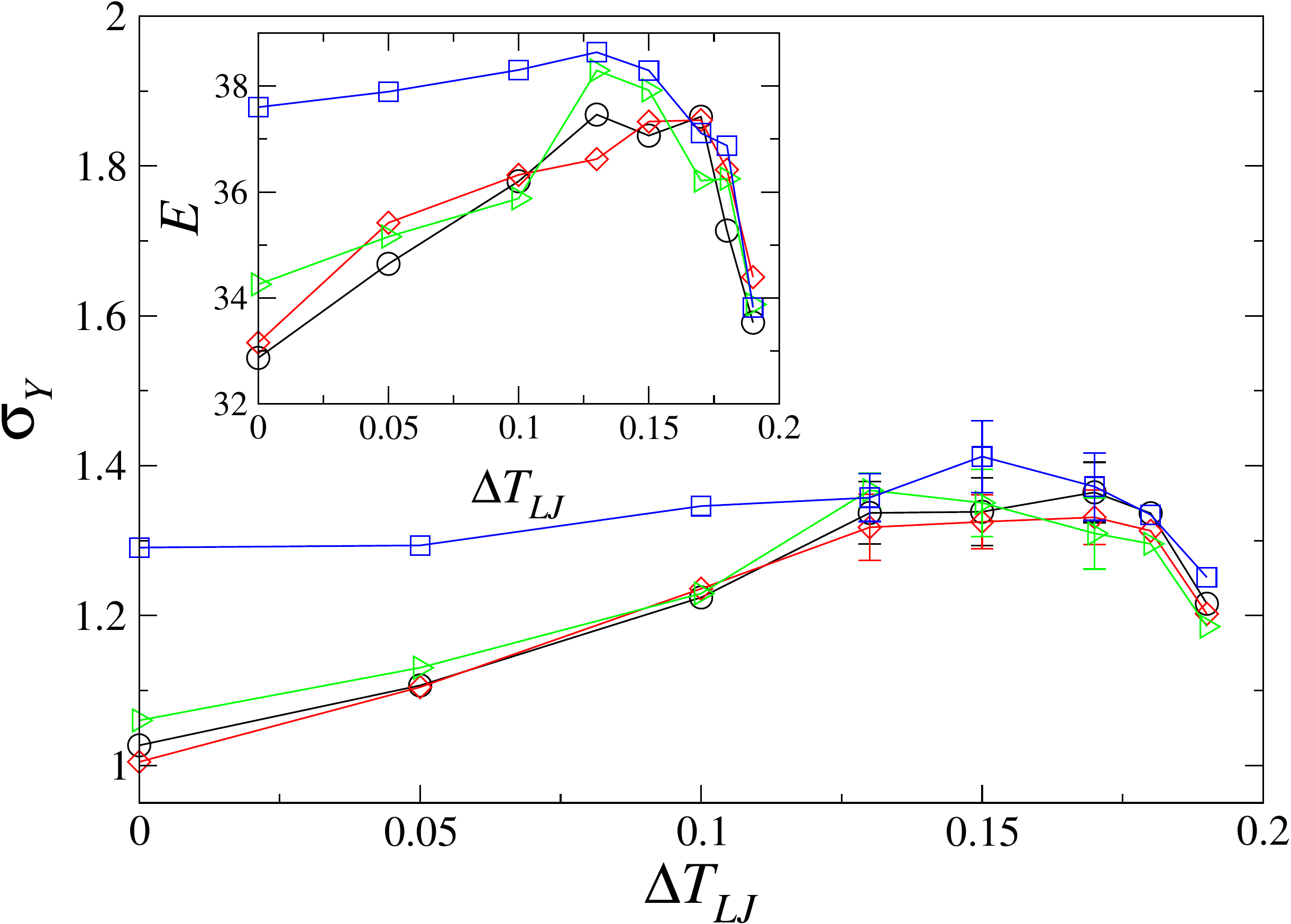}
\caption{(Color online) The dependence of the stress overshoot
$\sigma_Y$ (in units of $\varepsilon\sigma^{-3}$) as a function of
the thermal amplitude for glasses annealed with cooling rates
$10^{-2}\varepsilon/k_{B}\tau$ (black),
$10^{-3}\varepsilon/k_{B}\tau$ (red), $10^{-4}\varepsilon/k_{B}\tau$
(green), and $10^{-5}\varepsilon/k_{B}\tau$ (blue). The variation of
the elastic modulus $E$ (in units of $\varepsilon\sigma^{-3}$)
versus thermal amplitude is shown in the inset for the same samples.
}
\label{fig:yield_stress_E}
\end{figure}

The stress-strain curves reported in Fig.\,\ref{fig:stress_strain}
were used to extract the values of the yielding peak, $\sigma_Y$,
and the elastic modulus, $E$, which are plotted in
Fig.\,\ref{fig:yield_stress_E} as a function of the thermal
amplitude. It can be seen that $\sigma_Y$ increases with $\Delta
T_{LJ}$ and it has a maximum at $\Delta T_{LJ} \approx
0.15\,\varepsilon/k_B$ for all cooling rates. Note that the data are
somewhat scattered as simulations were performed for only one sample
in each case due to computational limitations. We also comment that
the yield stress only weakly depends on the cooling rate except for
the case $10^{-5}\varepsilon/k_{B}\tau$.  Similar trends can be
observed for the dependence of the elastic modulus on the thermal
amplitude and cooling rate, as illustrated in the inset to
Fig.\,\ref{fig:yield_stress_E}.  Generally, we find an inverse
correlation between the dependencies of the yielding peak in
Fig.\,\ref{fig:yield_stress_E} and $U(100\,T)$ in
Fig.\,\ref{fig:U100_delT} as functions of the thermal amplitude
$\Delta T_{LJ}$. In other words, the lower the energy state, the
higher the values of $\sigma_Y$ and $E$.  Finally, in comparison
with the results of the previous study, where thermal cycling was
performed with respect to a much lower reference temperature of
$0.01\,\varepsilon/k_B$~\cite{Priez18tcyc}, the relative increase in
$\sigma_Y$ and, correspondingly, decrease in $U(100\,T)$ for rapidly
cooled glasses is less pronounced for cycling with respect to the
reference temperature of $0.2\,\varepsilon/k_B$, considered in the
present study.

\section{Conclusions}

In summary, the response of amorphous alloys during the sequence of
quenching, periodic thermal treatment, and tensile loading was
investigated using molecular dynamics simulations.  The amorphous
material was represented via a binary mixture of atoms with highly
non-additive cross interactions that prevents crystallization upon
cooling.  The thermal quenching was performed at constant pressure
by cooling the binary mixture from the liquid state into the glassy
region with different rates. The reference temperature was chosen to
be approximately half the glass transition temperature, and the
simulations were performed at zero pressure.

\vskip 0.05in

After the thermal quench to different energy states, the glasses
were either set to age at the reference temperature or subjected to
one hundred thermal cycles of spatially uniform heating and cooling
at constant pressure.    It was found that thermal cycling leads to
relaxed states with the potential energy levels lower than those in
the aged samples for a given value of the cooling rate.   The
potential energy after one hundred cycles acquired a minimum at the
thermal amplitude just below the reference temperature.  The results
of uniaxial tensile loading demonstrated than the stress overshoot
and the elastic modulus only weakly depend on the cooling rate
except for the lowest rate. Overall, the inverse correlation between
the potential energy levels and mechanical properties for different
thermal amplitudes agrees well with the results of the previous
study~\cite{Priez18tcyc}, although the magnitude of the effects are
slightly reduced due to the higher reference temperature considered
in the present study.

\section*{Acknowledgments}

Financial support from the National Science Foundation (CNS-1531923)
is gratefully acknowledged. The article was prepared within the
framework of the Basic Research Program at the National Research
University Higher School of Economics (HSE) and supported within the
framework of a subsidy by the Russian Academic Excellence Project
`5-100'. The molecular dynamics simulations were performed using the
LAMMPS numerical code developed at Sandia National
Laboratories~\cite{Lammps}. Computational work in support of this
research was performed at Wright State University's Computing
Facility, the Ohio Supercomputer Center, and the HPC cluster at
Skoltech.

%%%%%%%%%%%%%%% FIGURES %%%%%%%%%%%%%%%%%%%%%%%

% snapshot of the system
%

% potential energy for cooling rates 10^-2, 10^-3, 10^-4, 10^-5 T=0.2
% inset densities
%

% temperature control over 5 periods
%

% potential energy for sample with cooling rate 10^-2
%
%

% potential energy for sample with cooling rate 10^-3
%
%

% potential energy for sample with cooling rate 10^-4
%
%

% potential energy for sample with cooling rate 10^-5
%
%

% U_100 versus delT
%

% PDF displacements at T_LJ=0.1 different cycles
%

% PDF displacements at different T_LJ
%

% stress-strain curves
%

% yield stress and elastic modulus vs amplitude of thermal cycling
%

\bibliographystyle{prsty}

\end{document}